\documentclass{osa-article}

\journal{osajournal}

\usepackage{tabulary}
\usepackage[per-mode=symbol, separate-uncertainty = true]{siunitx}
\usepackage{braket}
\usepackage{upgreek}

\begin{document}

\title{Performance of an optical single-sideband laser system for atom interferometry}

\author{Clemens Rammeloo\authormark{1}, Lingxiao Zhu\authormark{1,2}, Yu-Hung Lien\authormark{1,$\ast$}, Kai Bongs\authormark{1}, Michael Holynski\authormark{1,$\dag$}}

\address{\authormark{1}School of Physics and Astronomy, University of Birmingham, Edgbaston, Birmingham, B15 2TT, UK\\
         \authormark{2}College of Intelligence Science and Technology, National University of Defense Technology, Changsha, 410073, China}

\email{\authormark{$\ast$}y.lien@bham.ac.uk}
\email{\authormark{$\dag$}m.holynski@bham.ac.uk}

\begin{abstract*}
This paper reports on a detailed performance characterization of a recently developed optical single-sideband (OSSB) laser system based on an IQ modulator and second-harmonic generation for rubidium atom interferometry experiments. The measured performance is used to evaluate the noise contributions of this OSSB laser system when it is applied to drive stimulated Raman transitions in $^{87}$Rb for precision measurements of gravitational acceleration. The laser system suppresses unwanted sideband components, but additional phase shift compensation needs to be applied when performing frequency chirps with such an OSSB laser system. The total phase noise contribution of the OSSB laser system in the current experiment is {72~mrad} for a single atom-interferometry sequence with interrogation times of {$T=120$~{ms}}, which corresponds to a relative precision of {32~{n}$g$} per shot. The dominant noise sources are found in the relative intensity fluctuations between sideband and carrier components and the phase noise of the microwave source.
\end{abstract*}

\section{Introduction}

Atom interferometry is increasingly being applied outside laboratory environments, for instance, in surveys of precision measurements of gravitational acceleration \cite{Farah2014, Wu2014, Freier2015, Menoret2018} and on mobile platforms \cite{Barrett2016, Bidel2018, Becker2018}. This drives a demand for compact and efficient laser systems that address transitions in alkali atoms. Such precision measurements rely on two laser beams having both a stable phase-relation and a controlled frequency difference, for which several compact laser systems have been developed. These laser systems can generally be categorized in either consisting of two lasers that are phase stabilized with respect to each other \cite{Schmidt2010, Wu2014} or a single laser with a phase modulation to generate multiple frequency components \cite{Carraz2009, Theron2014, Wu2017, Diboune2017, Li2017a, Fang2018}. The latter requires only a single seed-laser, thus reducing complexity and potentially enabling more compact designs. However, most laser systems that are based on direct phase modulation apply an electro-optic modulator (EOM) which also generates undesired frequency components. This method is relatively inefficient as it wastes the available optical power and could drive parasitic transitions that perturb measurements of gravitational acceleration \cite{Carraz2012, Zhu2018OSSB}.

A new method that suppresses undesirable sideband components is based on an IQ modulator (IQM) that generates an optical single-sideband (OSSB) for tunable laser frequency components with a single seed-laser \cite{Zhu2018OSSB}. This laser system design relies on components from the telecommunication industry operating at \SI{1560}{\nano\meter} wavelengths and second-harmonic generation to \SI{780}{\nano\meter} wavelengths in order to drive transitions in Rb atoms. We have previously demonstrated the applicability of an OSSB laser system in measurements of gravitational acceleration \cite{Zhu2018OSSB}, including the suppression of systematic phase shifts from undesirable sidebands. In general, such an IQM-based laser system can be used in all stages of a typical atom-interferometry experiment that include laser cooling and detection of rubidium atoms \cite{Gouraud2019}. The application of an IQM instead of an EOM provides additional controls over the sideband to carrier ratios. Due to this, however, an OSSB laser system has the potential to introduce significant intensity changes in sideband and carrier components.

In this work we discuss a detailed performance characterization of an OSSB laser system, with a focus on the purpose of driving stimulated Raman transitions between the hyperfine-split ground-states in $^{87}$Rb. These transitions are applied in Mach{\textendash}Zehnder (MZ) type pulse sequences for precision measurements of gravitational acceleration \cite{Kasevich1991}. Here we evaluate for the parameters of our atom-interferometry experiment, the impact of the noise contributions from the OSSB laser system. This paper first describes the OSSB laser system setup and its wavelength conversion performance. This is followed by measurements of the output power, intensity noise and stability of the sideband component. We also identify additional phase shifts that need to be considered when doing frequency chirps with an IQM. Finally, we report on laser phase and frequency noise measurements, and compare the expected phase shift contributions from all these effects for our MZ type atom-interferometer.

\section{Laser system setup}\label{sec:Setup}

\begin{figure}[htbp]
\centering
\includegraphics[width=0.85\linewidth]{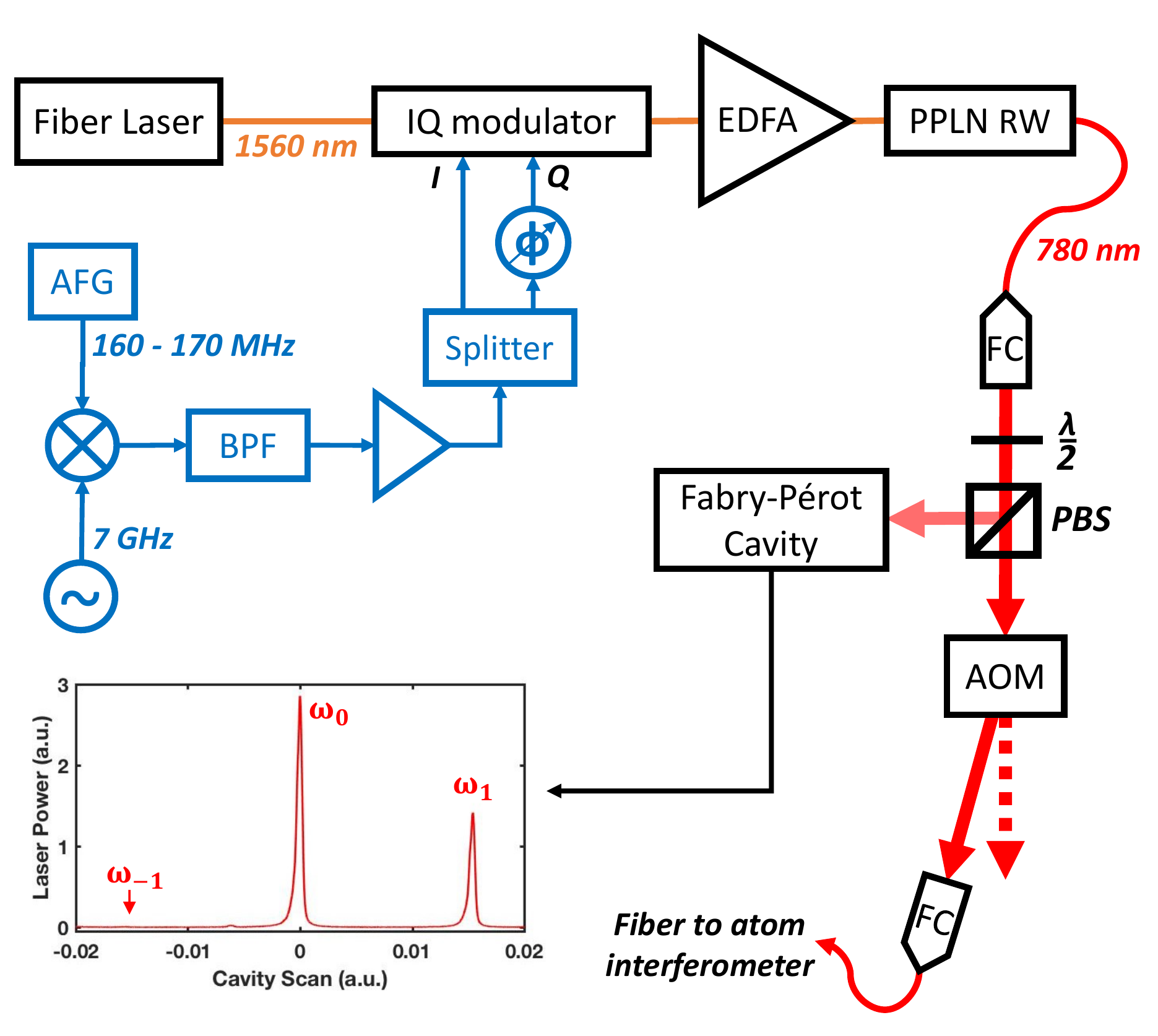}
\caption{Diagram of the optical single-sideband laser system, see text for details. EDFA: erbium-doped fiber amplifier, PPLN RW: periodically-poled lithium niobate ridge-waveguide, AFG: arbitrary function generator, BPF: bandpass filter, FC: fiber coupler, PBS: polarizing beamsplitter, AOM: acousto-optic modulator. The plot is a measured transmission spectrum of a Fabry{\textendash}P\'erot cavity scan showing the laser spectrum after the PPLN RW. The carrier $\omega_{0}$ and first-order sideband $\omega_{1}$ are the laser frequency components used to drive the two-photon Raman transition, while the other $\omega_{-1}$ sideband is suppressed through tuning of the RF phase shift and bias voltages on the IQ modulator.}
\label{fig:LaserSystem}
\end{figure}

The OSSB laser system applied in this work is based on second harmonic generation (SHG) of light from components operating at \SI{1560}{\nano\meter} wavelengths as shown in the diagram in {Fig.~\ref{fig:LaserSystem}}. The use of fiber-coupled components from the C-band telecommunication industry enables a relatively efficient and compact design. Additionally, to the best of our knowledge there are currently no IQ modulators commercially available for wavelengths outside telecommunication bands.

The seed-laser is an erbium-doped fiber laser (NKT Photonics, Koheras BASIK E15) and the single-sideband frequency component is generated via an IQ modulator (Photline, MXIQ-LN-40). An erbium-doped fiber amplifier (Orion Laser, YEDFA-PM) boosts the power to about \SI{1.5}{\watt} before an SHG module (NTT Electronics, WH-0780-000-F-B-C) which converts the light to \SI{780}{\nano\meter} wavelengths. The latter consists of a periodically-poled lithium niobate ridge-waveguide (PPLN RW) that is temperature stabilized and has its input and output ports pigtailed with polarization maintaining (PM) fibers. The output from the SHG fiber is collimated and passed through a polarizing beam splitter (PBS) to filter polarization fluctuations. A half-wave plate before the PBS is set such that about \SI{1}{\percent} of the light is reflected by the PBS towards a scanning Fabry{\textendash}P\'erot cavity for recording the relative powers between the laser frequency components. The transmitted beam after the PBS is switched by an acousto-optic modulator (AOM) to create the pulse sequences for the Raman beam. The first-order diffracted output from the AOM is coupled into a PM fiber to the Raman beam collimator of the atom interferometer. Details of the atom-interferometer experiment can be found in \cite{Zhu2018Thesis}.

The laser frequency components that drive the two-photon Raman transition are the carrier $\omega_{0}$ and +1st order sideband $\omega_{1}$ generated by the IQ modulator and subsequent wavelength conversion. The tunable RF signal for the IQ modulation is generated by mixing a \SI{7}{\giga\hertz} signal from a signal generator (Keysight, E8267D) with the output of an arbitrary function generator (Tektronix, AFG3252C) via an RF mixer (Mini-Circuits, ZMX-10G+). The down converted frequency component is around \SI{6.835}{\giga\hertz}, to be resonant with the hyperfine-splitting frequency of the ${^5S}_{1/2}$ state in $^{87}$Rb, while other RF components are suppressed by more than \SI{50}{\decibel} by a cavity bandpass filter (ELHYTE, BP6834-70/T-5CS). After amplification (Nextec, NBL00426), the RF signal is split for the I and Q ports. A tunable RF phase shift is introduced before the Q port by either a manual phase shifter (Fairview Microwave, SMP0820) or a voltage controlled phase shifter (RF-Lambda, RVPT0408GBC).

The optical output spectrum of the OSSB laser system shown in {Fig.~\ref{fig:LaserSystem}}, is measured by the scanning Fabry{\textendash}P\'erot cavity. These recordings are used to monitor and to set the +1st order sideband / carrier power ratio to ${\sim}{1/2}$ such that the first-order AC Stark shifts are canceled in our experiment. The -1st order sideband $\omega_{-1}$ is suppressed below \SI{-23}{\decibel} with respect to the carrier by tuning the RF phase shifter and bias voltages on the IQ modulator \cite{Zhu2018OSSB}.

\section{Wavelength conversion and output power}

For an efficient second-harmonic generation of \SI{780}{\nano\meter} wavelength light the quasi-phase-matching condition in the PPLN RW needs to be maintained, which depends on both the laser frequency and the temperature of the PPLN RW \cite{Hum2007}. The frequency bandwidth of the quasi-phase-matching condition is measured by tuning the fiber laser frequency and simultaneously recording both frequency and power of the \SI{780}{\nano\meter} wavelength light at the PPLN RW output. The thus measured FWHM bandwidth is \SI{80(1)}{\giga\hertz} and is much larger than both the sideband separation as well as the frequency tuning range required in our atom-interferometry experiment.

\begin{figure}[htbp]
\centering
\includegraphics[width=0.85\linewidth]{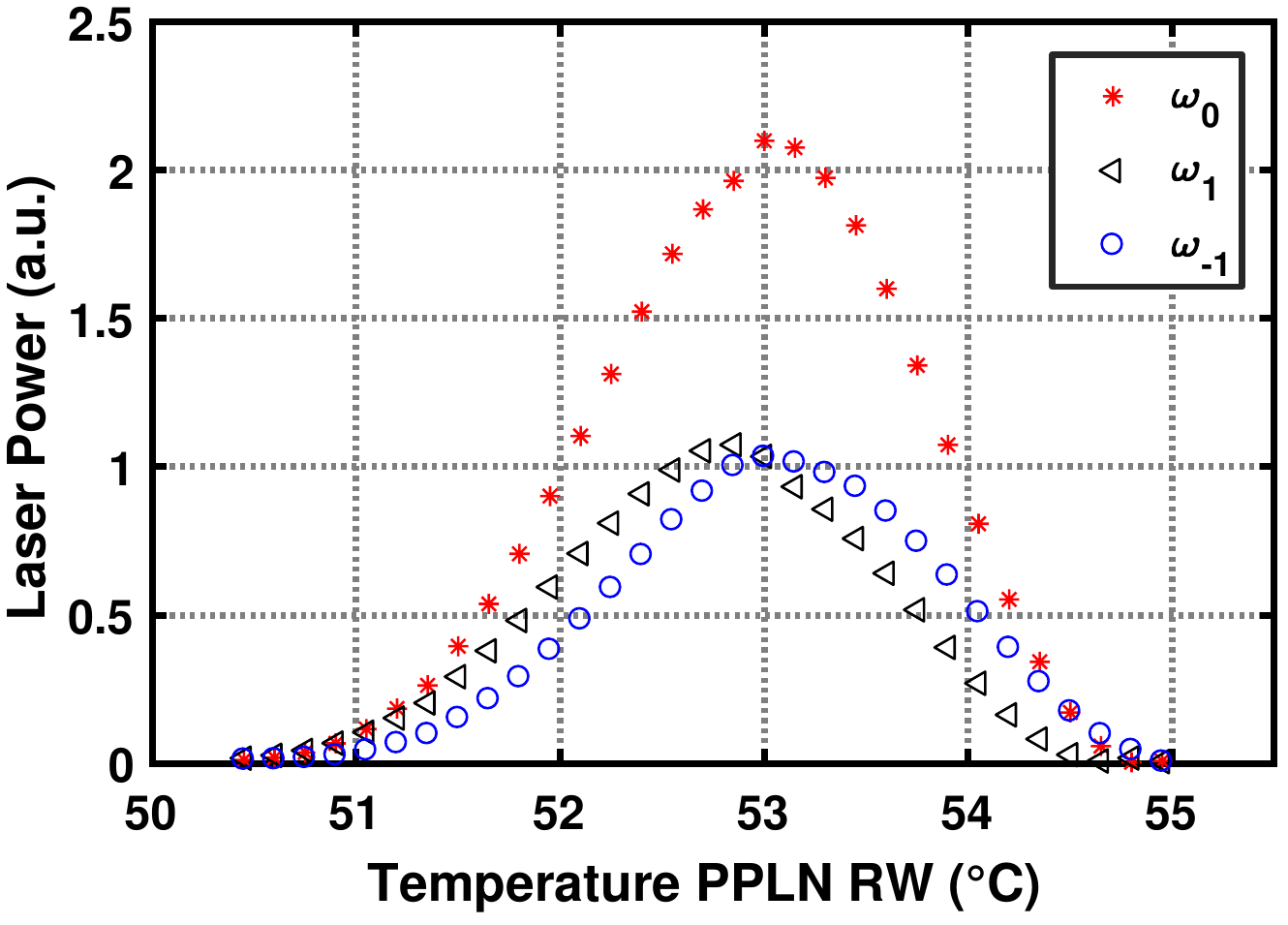}
\caption{SHG output power of the carrier $\omega_{0}$ and $\pm 1$st order sidebands $\omega_{\pm 1}$ at \SI{780}{\nano\meter} as a function of the temperature of the PPLN RW. The $\omega_{-1}$ sideband is not suppressed in this measurement to show the shift in the optimum PPLN RW temperature.}
\label{fig:PPLNTemperature}
\end{figure}

The temperature for optimum SHG is around \SI{53}{\celsius} for the particular PPLN RW used in this laser system as shown in {Fig.~\ref{fig:PPLNTemperature}}. The PPLN RW temperature is stabilized by a compact temperature controller (Meerstetter Engineering, TEC-1091) that keeps the temperature fluctuations within ${\pm\SI{0.005}{\celsius}}$. The temperature range over which the quasi-phase-matching condition of this PPLN RW is maintained has a HWHM of \SI{0.9}{\celsius} around the optimum temperature. However, the optimum phase-matching temperatures of the sideband frequency components are offset from the carrier's optimum by about \SI{0.15}{\celsius} as seen in {Fig.~\ref{fig:PPLNTemperature}}. This offset can be exploited for increased sideband suppression. For instance, if the PPLN RW temperature is tuned to a slightly lower temperature, the phase-matching condition for the $\omega_{1}$ sideband is enhanced while simultaneously reducing that of the $\omega_{-1}$ sideband. This can result in an additional \SI{2}{\decibel} in sideband suppression in the \SI{780}{\nano\meter} wavelength output, but this method has a trade-off with a reduction in total output power.

\begin{figure}[htbp]
\centering
\includegraphics[width=0.85\linewidth]{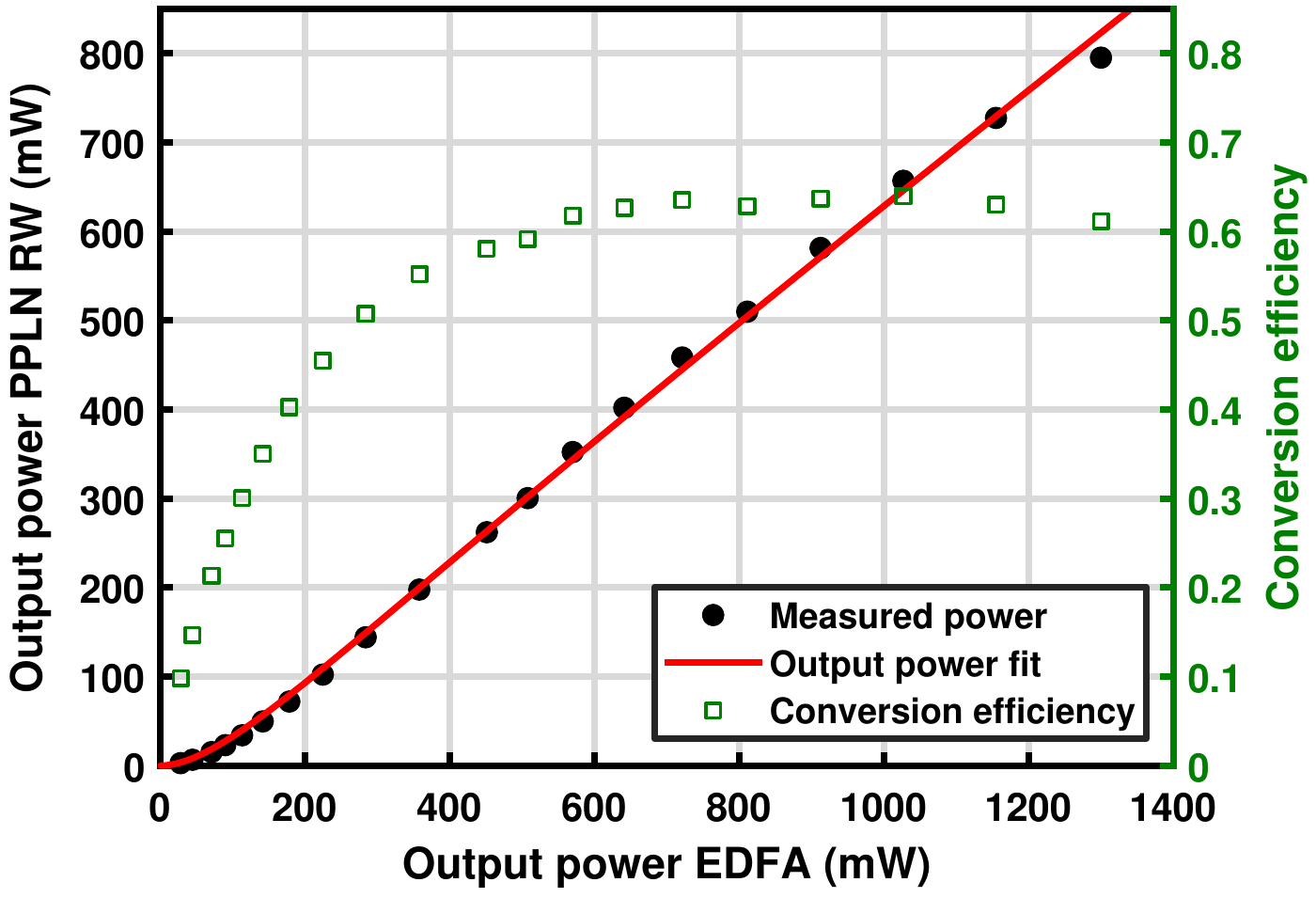}
\caption{The optical power at the output of the PPLN RW (\SI{780}{\nano\meter} wavelength) as a function of the EDFA output (\SI{1560}{\nano\meter} wavelength) and the conversion efficiency as the ratio between both powers. From the fit with \eqref{equ:SHGPowerConversion} a second-harmonic generation efficiency of $\eta=\SI{1.3(3)e3}{\percent\per\watt}$ and a coupling efficiency of $\epsilon=\SI{64(1)}{\percent}$ are extracted.}
\label{fig:PowerConversion}
\end{figure}

The wavelength conversion efficiency is characterized by measuring the \SI{780}{\nano\meter} wavelength power after the PPLN RW output fiber as a function of the \SI{1560}{\nano\meter} wavelength power supplied by the EDFA, see {Fig.~\ref{fig:PowerConversion}}. The relation between the output power of a PPLN medium $P_{2\omega}$ and the input power $P_{\omega}$ is described by the equation \cite{Chiodo2013}
\begin{equation}
	\label{equ:SHGPowerConversion}
	P_{2\omega}=\epsilon P_{\omega}\tanh^{2}{\!\left(\sqrt{\eta\epsilon P_{\omega}}\right)}.
\end{equation}
Here is $\eta$ the SHG efficiency and $\epsilon$ the light coupling efficiency. In our laser system, $\epsilon$ represents the accumulated coupling efficiencies between the ridge-waveguide and its input and output fibers, as well as the fiber-to-fiber coupling efficiency from the EDFA to the SHG module.

In the results of {Fig.~\ref{fig:PowerConversion}} an almost linear relation is observed for optical powers above about \SI{200}{\milli\watt} where the wavelength conversion is limited by the light coupling efficiency. Fitting the measurements with \eqref{equ:SHGPowerConversion} shows a coupling efficiency of $\epsilon=\SI{64(1)}{\percent}$ and an SHG efficiency $\eta=\SI{1.3(3)e3}{\percent\per\watt}$. The SHG efficiency $\eta$ is higher compared to bulk PPLN crystals \cite{Wang2016, Theron2017}, but the output power of these PPLN RW modules reaches saturation for input powers around \SI{2}{\watt} \cite{Leveque2014}. The output power of the OSSB laser system presented here, enables a $\pi/2$-pulse duration of $\tau_\mathrm{R}=\SI{25}{\micro\second}$ for the current Raman beam size in our atom-interferometry experiment at a red-detuning of \SI{2}{\giga\hertz} with respect to the ${^5S}_{1/2}\leftrightarrow {^5P}_{3/2}$ transition in $^{87}$Rb.

\section{Intensity noise and sideband/carrier ratio}

Since the OSSB laser system is applied to drive stimulated Raman transitions between atomic ground states, laser power fluctuations cause variations in the Rabi frequency and thus reduce the signal to noise ratio in the measurements of the atomic state population \cite{Peters2001}. Besides this effect, the influence of short-term intensity noise during an interferometry sequence can cause phase shifts at the atom-interferometer output. These phase shifts are usually suppressed through tuning of the sideband/carrier intensity ratio. However, it will be shown that an IQM-based laser system can experience significant variations in the sideband/carrier ratio. In the following sections we investigate the impact from the measured laser power fluctuations and the changes in the sideband/carrier ratio.

\subsection{Power stability and RIN}

\begin{figure}[htbp]
\centering
\includegraphics[width=0.85\linewidth]{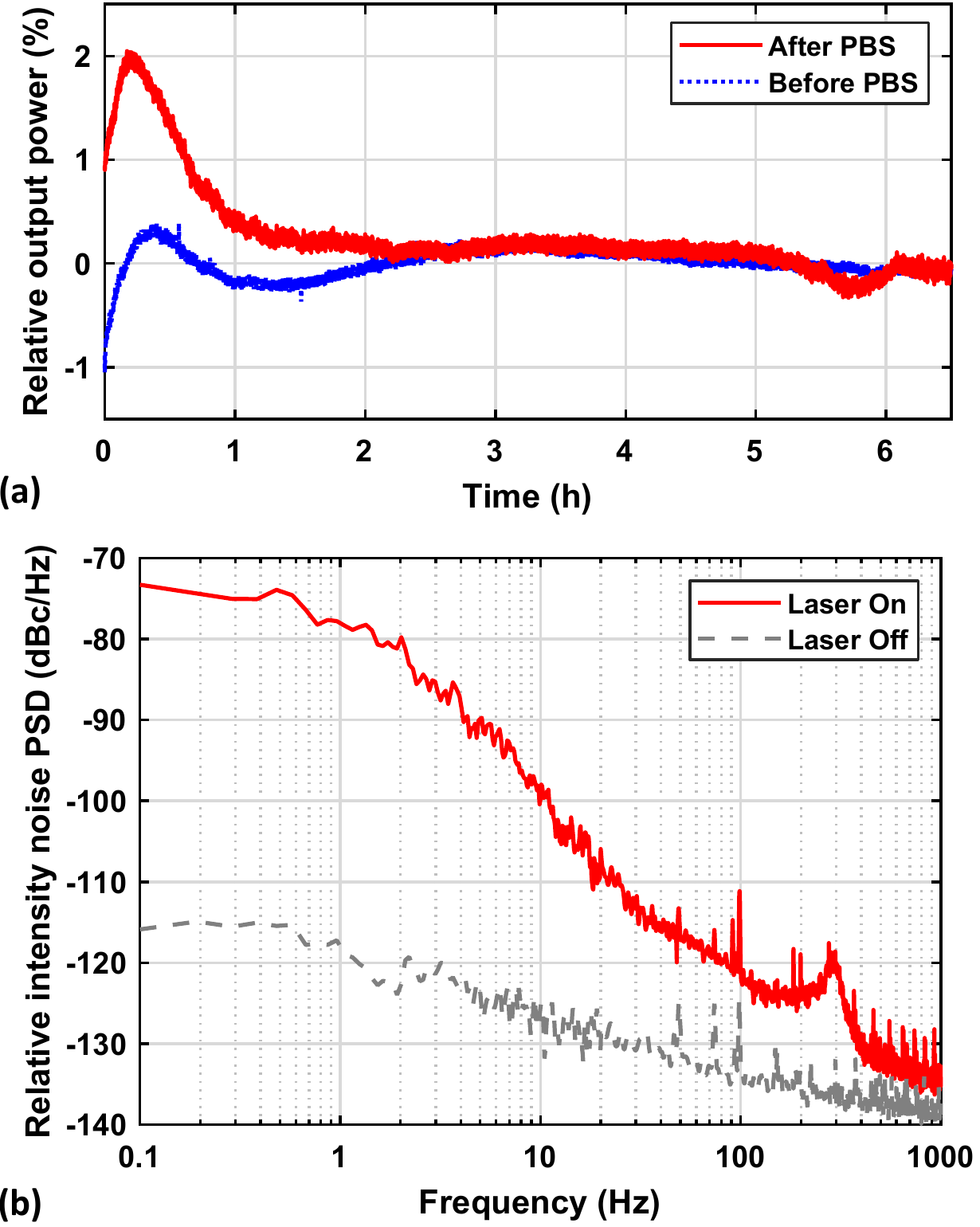}
\caption{\textbf{(a)} Relative power of the \SI{780}{\nano\meter} wavelength output of the laser system with respect to an average output power of \SI{0.4}{\watt} as recorded before and after the polarizing beam splitter (PBS). \textbf{(b)} Power spectral density (PSD) of the relative intensity noise measured at the output fiber (solid line) with respect to the measurement noise floor (dashed line). The PSD is averaged for frequencies above \SI{20}{\hertz} for clarity.}
\label{fig:RelativePowerRIN}
\end{figure}

Measurements of both the long-term stability of the laser system's output power over several hours, as well as the short-term relative intensity noise (RIN) spectrum are presented in {Fig.~\ref{fig:RelativePowerRIN}}. At the start of the output power recording in {Fig.~\ref{fig:RelativePowerRIN}a} a peak is observed during the warm-up period of the laser system, but after an hour the long-term relative drift in output power reduces to below \SI{0.5}{\percent}. When comparing the power fluctuations before and after the PBS, it is seen that one of the main causes of the power drifts originate from the changing polarization at the output fiber of the PPLN RW.

The short-term power fluctuations are measured with a \SI{125}{\mega\hertz}-bandwidth photodiode (TTi, TIA-525) connected to a spectrum analyzer (Tektronix, RSA5115B). This allows the recording of the RIN spectrum of the laser system, shown in {Fig.~\ref{fig:RelativePowerRIN}b}. When integrating the RIN spectrum over the sensitive frequency range of a MZ pulse sequence with $\tau_\mathrm{R}=\SI{25}{\micro\second}$ and an interrogation time $T=\SI{120}{\milli\second}$ between the pulses, the integrated RIN is estimated to be \SI{-65}{\decibel}c for a single MZ pulse sequence.

\begin{figure}[htbp]
\centering
\includegraphics[width=0.85\linewidth]{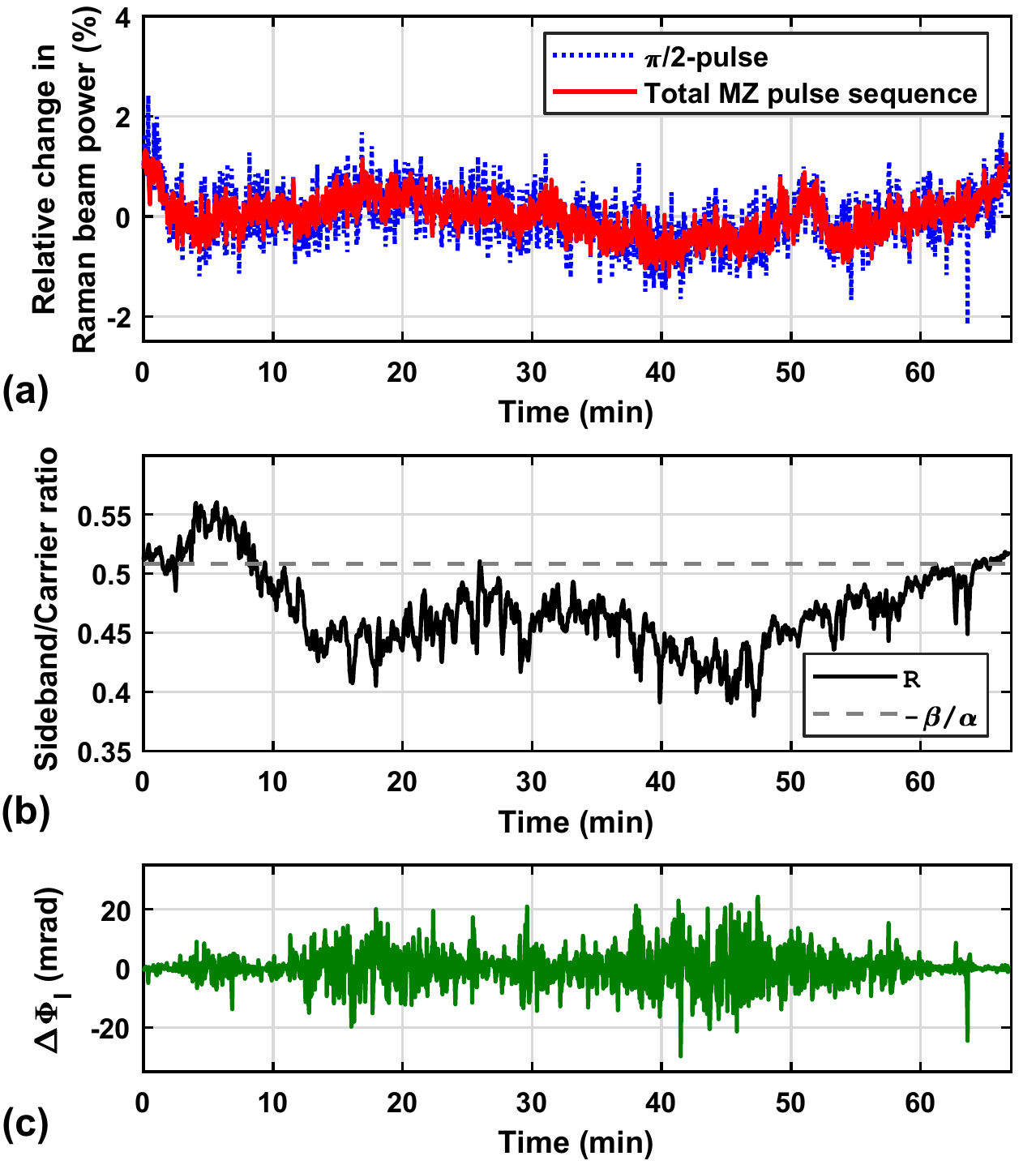}
\caption{Simultaneous measurements of \textbf{(a)} the relative change in total beam power for repeated Mach{\textendash}Zehnder pulse sequences in comparison with single $\pi/2$-pulses, and \textbf{(b)} the sideband/carrier ratio $R$ in comparison to the ratio $-\beta/\alpha$ where the light shift is canceled. \textbf{(c)} The phase shift $\Delta\Phi_I$ in the atom interferometer calculated via \eqref{equ:PhaseShift_IntensityNoise} with the measured intensity and ratio $R$.}
\label{fig:PulseIntensityRatioShift}
\end{figure}

To evaluate the effect of these power fluctuations on the atom interferometer, repeated MZ pulse sequences are produced with the AOM while the pulse powers are measured with the high-bandwidth photodiode at the output fiber of the OSSB laser system. Each pulse sequence is integrated and normalized to the average total Raman beam power to create the plot in {Fig.~\ref{fig:PulseIntensityRatioShift}a}. The variation in the integrated Raman beam power over a complete MZ pulse sequence is smaller in comparison to a single $\pi/2$-pulse, due to the averaging over the three pulses. From the relative power fluctuations in {Fig.~\ref{fig:PulseIntensityRatioShift}a} we estimate its contribution to the atom-interferometer signal-to-noise ratio to about \num{50/1}. This effect could in principle be mitigated through power stabilization with the AOM, or reduced in post-process when the Raman beam power is monitored for each MZ pulse sequence.

\subsection{Sideband/carrier ratio}\label{sec:SidebandCarrierRatio}

Intensity fluctuations in the Raman laser beams during the pulse sequence cause varying light shifts on the atomic states\cite{LeGouet2008}. The differential light shift contribute to phase shifts in the atom interferometer, but can be suppressed through optimization of the intensity ratio between sideband and carrier components. However, the sideband/carrier ratio in the OSSB laser system exhibits a drift due to instabilities in the IQ modulator biases \cite{Salvestrini2011}. This is clearly observed in the measurements from the scanning Fabry{\textendash}P\'erot cavity plotted in {Fig.~\ref{fig:PulseIntensityRatioShift}b}. These measurements were performed simultaneously with the total power fluctuations recorded for {Fig.~\ref{fig:PulseIntensityRatioShift}a}. This enables us to evaluate the effects on the atom-interferometer phase shift from the varying sideband/carrier ratio both independently from, and in combination with the measured intensity noise. The phase shift contribution from the resulting differential light shift could be reduced through the application of an active stabilization of the IQ modulator biases, or via the subtraction of the phase shift obtained with the following method in post-process.

As each of the Raman laser beams are detuned with respect to the electronic transition ${^{5}S_{1/2}}\leftrightarrow {^{5}P_{3/2}}$ in $^{87}$Rb, they induce a light shift on the two-photon Raman transition. The detuning $\delta\nu$ due to this light shift can be expressed as a linear combination of the laser intensities \cite{Bateman2010}: $\delta\nu=\alpha I_{1}+\beta I_{0}$. Here are $I_{1}$ and $I_{0}$ the intensities of the +1st-order sideband component $\omega_{1}$ and the carrier $\omega_{0}$, respectively. The $\alpha$ and $\beta$ components are calculated using the formalism from \cite{Cheinet2006PhD} and for Raman beams with a red-detuning of \SI{2}{\giga\hertz}, we find $\alpha=\SI{5.43}{\kilo\hertz\per\milli\watt\centi\meter\squared}$ and $\beta=-\SI{2.76}{\kilo\hertz\per\milli\watt\centi\meter\squared}$. Because we measure the ratio $R=I_{1}/I_{0}$ with the scanning Fabry{\textendash}P\'erot cavity and the total Raman beam intensity $I_\mathrm{tot}=I_{1} + I_{0}$ with a fast photodiode, the detuning from the differential light shift is expressed here as
\begin{equation}
	\label{equ:LightShift}
	\delta\nu = \alpha I_\mathrm{tot} \left( \frac{R+\beta/\alpha}{R+1} \right) .
\end{equation}

This light shift is canceled by controlling the intensity ratio between sideband and carrier such that $R=-\beta/\alpha$, which is in our experiment approximately \num{0.51}. In the current laser system, however, the ratio $R$ drifts in the range of \SIrange{15}{25}{\percent} relative to the $-\beta/\alpha$ value as observed in {Fig.~\ref{fig:PulseIntensityRatioShift}b}. Depending on $R$, the stimulated Raman transitions become sensitive to temporal fluctuations in intensity during the interferometry sequence, which leads to a phase shift $\Delta\Phi_I$ in the atom interferometer given by \cite{LeGouet2008}
\begin{equation}
	\label{equ:PhaseShift_IntensityNoise}
	\Delta\Phi_I = \int_{-\infty}^{+\infty}g(t)2\pi\delta\nu(t)\mathrm{d}t .
\end{equation}
In this equation is $g(t)$ the sensitivity function of a MZ type pulse sequence as derived by \cite{Cheinet2008}.

The expected phase shift $\Delta\Phi_I$ with the current OSSB laser system is determined in two cases. First, the contribution from the fluctuating total intensity noise $I_\mathrm{tot}(t)$ is evaluated while assuming a constant ratio $R$ in \eqref{equ:LightShift} during each MZ pulse sequence. Secondly, the effect from a varying sideband/carrier ratio $R(t)$ during the pulse sequence is determined by setting the total intensity $I_\mathrm{tot}$ equal to the average intensity of the Raman laser beams.

The phase noise contribution from the total intensity noise is evaluated by performing the integration of \eqref{equ:PhaseShift_IntensityNoise} over each measured pulse sequence. Fast changes in the sideband/carrier ratio during an interferometry sequence cannot be measured with the scanning Fabry{\textendash}P\'erot cavity due to its limited scan speed. Therefore, the ratio $R$ is assumed constant during a MZ single sequence when evaluating the integration of \eqref{equ:PhaseShift_IntensityNoise} over the measured intensity $I_\mathrm{tot}(t)$. The resulting phase shifts for each pulse sequence are plotted in {Fig.~\ref{fig:PulseIntensityRatioShift}c} and reach about \SI{30}{\milli\radian} per shot with a standard deviation of \SI{6}{\milli\radian} over an hour of repeated pulse sequences. These phase shifts are higher than those reported in other works \cite{LeGouet2008}, but can be subtracted in post-process. However, this method is currently limited by the measurement precision of the ratio $R$.

Fast changes of the carrier and sideband ratio $R(t)$ on time scales down to \SI{10}{\milli\second} are measured via a beat signal spectrum generated in combination with a second laser operating at \SI{780}{\nano\meter} wavelengths. A spectrogram of the beat signal with both carrier and sideband components of the OSSB laser system is recorded by a real-time spectrum analyzer (Tektronix, RSA5115). The sideband/carrier ratio in these measurements shows RMS fluctuations below \SI{0.9}{\percent} on time-scales of a single interferometry sequence. We evaluate the effect of this changing sideband/carrier ratio $R(t)$ during a MZ pulse sequence again via \eqref{equ:LightShift} and \eqref{equ:PhaseShift_IntensityNoise}. For our atom-interferometry experiment with an average intensity $I_\mathrm{tot} = \SI{34}{\milli\watt\per\centi\meter\squared}$, the light shift sensitivity is \SI{12}{\kilo\hertz\per\percent} change in the sideband/carrier ratio. Thus the measured variation in the ratio $R$ between pulses corresponds here to a contribution of \SI{56}{\milli\radian} in the atom-interferometer phase shift. The effect on the determination of the gravitational acceleration $g$ using MZ pulse sequences with $T=\SI{120}{\milli\second}$ is then at a level of ${\num{25}~\mathrm{n}g}$ per shot.

The changing intensity $I_\mathrm{tot}(t)$ not only originate from the laser system's RIN, but also from the movement of the atoms in the Raman laser beam. The effect of the latter depends on the experimental conditions, for example the atom-cloud temperature, but can be suppressed via differential measurements with reversed frequency chirps \cite{Louchet-Chauvet2011}. Such methods do not, however, eliminate uncorrelated fluctuations in intensity or sideband/carrier ratio. Recording the sideband/carrier ratio during MZ pulse sequences, or stabilizing $R$ via the IQM bias voltages \cite{Li2017b}, is thus necessary when using this OSSB laser systems for precision measurements of gravity.

\section{Frequency chirping}

In order to keep the Raman laser on resonance with the accelerating atoms in the atom interferometer, the RF signal applied to the IQ modulator is chirped. These frequency chirps can introduce variations in the sideband/carrier ratio due to a changing RF phase difference between the I- and Q-ports. This frequency dependence is attributed to the microwave components and dispersion in the RF cables, the result of which could cause a bias in the atom-interferometer phase shift.

\begin{figure}[htbp]
\centering
\includegraphics[width=0.85\linewidth]{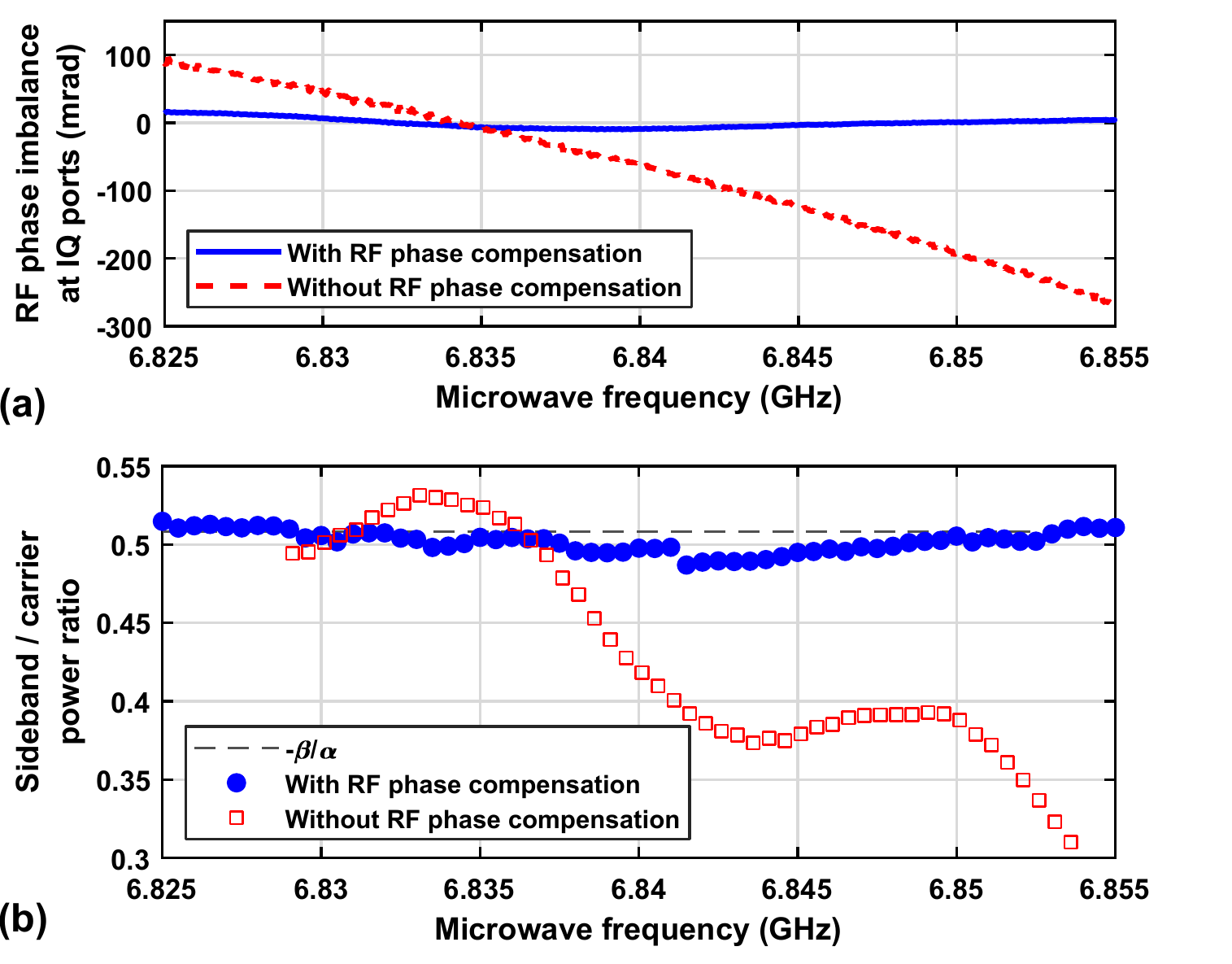}
\caption{The effect of a changing microwave frequency for the single-sideband generation on \textbf{(a)} the phase difference between the RF signals at the I- and Q-ports of the IQ modulator, and \textbf{(b)} the power ratio between +1st-order sideband and carrier components.}
\label{fig:FrequencyChirpPhaseShift}
\end{figure}

The RF phase imbalance is measured by combining the signals for the I- and Q-ports with an RF mixer (Mini-Circuits, ZMX-10G+) and recording the DC voltage output while chirping the RF signal. The resulting phase imbalance, within the microwave frequency range applicable in our atom-interferometer, is shown in {Fig.~\ref{fig:FrequencyChirpPhaseShift}a} for two different configurations. In case a manual RF phase shifter is used, there is no active compensation of the RF imbalance and we observe a variation of about \SI{11}{\milli\radian\per\mega\hertz}. On the other hand, when a voltage controlled phase shifter is applied to compensate the frequency dependent RF phase shifts, this variation is reduced below \SI{3}{\milli\radian\per\mega\hertz}.

The effect of this RF phase compensation on the power ratio between +1st-order sideband and carrier components is clearly visible in {Fig.~\ref{fig:FrequencyChirpPhaseShift}b}. The variation in the ratio $R$ as measured with the Fabry{\textendash}P\'erot cavity, is about \SI{4}{\percent} when the RF phase compensation is applied. However, this measurement is limited by the temporal variations in $R(t)$ which were already discussed in {section~\ref{sec:SidebandCarrierRatio}}, thus the frequency chirping effect on the sideband/carrier ratio is likely smaller. Without this RF phase shift compensation the changes in the ratio $R$ are an order of magnitude larger. It should be noted that with the applied RF phase compensation, the -1st-order sideband stays suppressed to a level \SI{-23}{\decibel} with respect to the carrier.

The contribution to the atom-interferometry phase shift can be considered as a bias when the same frequency chirp is applied during each measurement cycle. We estimate from the measurements of {Fig.~\ref{fig:FrequencyChirpPhaseShift}b} and \eqref{equ:PhaseShift_IntensityNoise}, that the RF phase compensation can reduce this frequency chirping bias to below \SI{82}{\milli\radian} for MZ pulse sequences with $T=\SI{120}{\milli\second}$. This bias could in principle be corrected either in post-process with a characterization of the microwave components and IQ modulator, or via feedback from the Fabry{\textendash}P\'erot cavity measurements to the voltage controlled phase shifter. Foremost, this effect highlights the need for careful RF design when applying IQM-based laser systems to perform accurate measurements using atom interferometry.

\section{Laser phase noise}

The main noise source in many atom-interferometers for precision measurements of gravitational acceleration, is the phase noise between the two Raman laser beams. The phase noise between two retro-reflected Raman beams is usually dominated by vibrations of the retro-reflecting mirror. These perturbing accelerations are filtered through vibration isolation \cite{Peters2001, Farah2014, Freier2015} or compensated using recordings from classical accelerometers \cite{Lautier2014b, Cheiney2018}. The remaining limit of the Raman beam phase noise is then governed by phase fluctuations between the two frequency components of the Raman laser system.

\begin{figure}[htbp]
\centering
\includegraphics[width=0.85\linewidth]{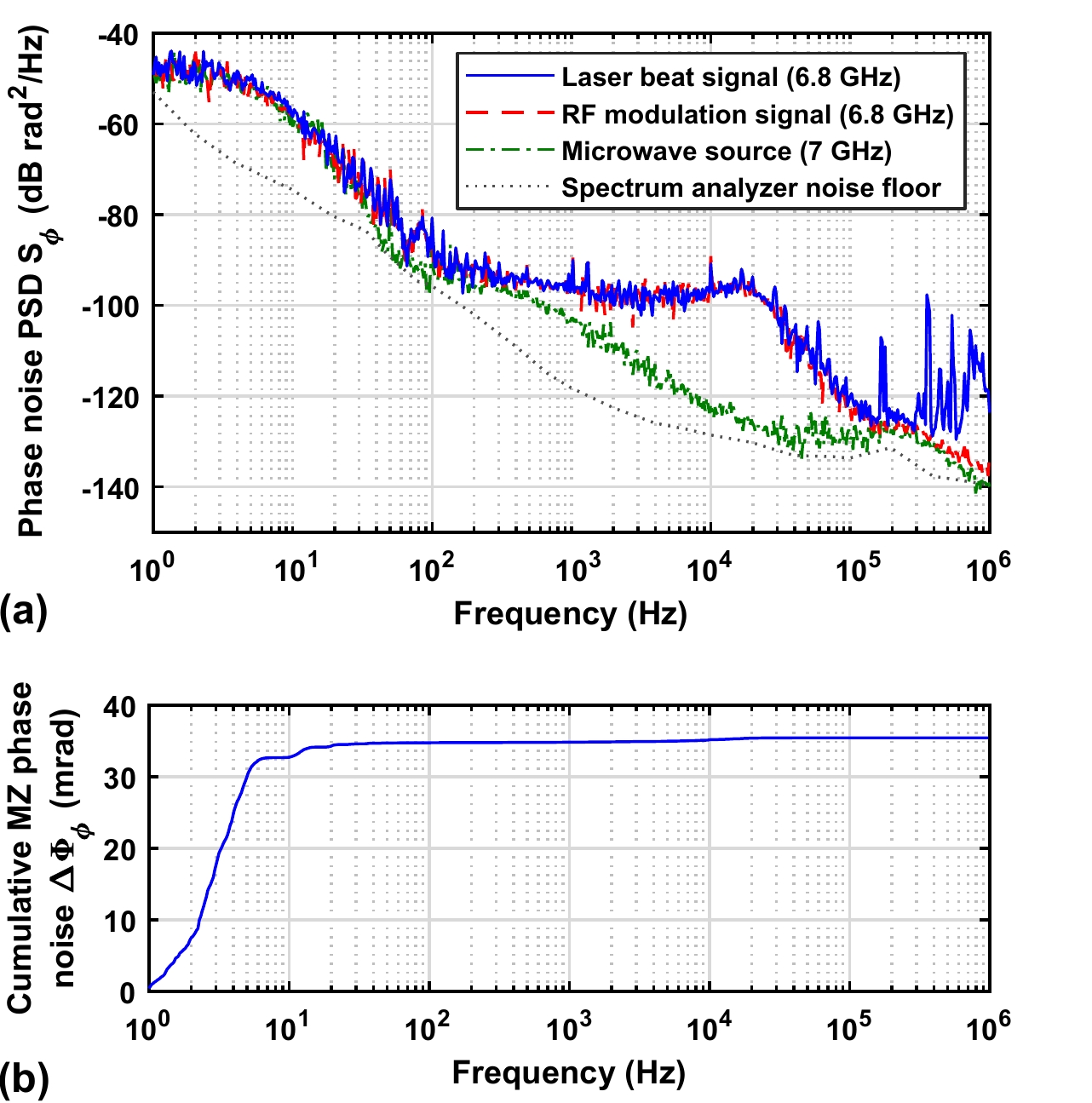}
\caption{\textbf{(a)} The power spectral density (PSD) of the phase noise between the Raman laser frequency components measured in the optical beat signal. The measured microwave signal used to create the sideband components and the \SI{7}{\giga\hertz} microwave source are also plotted in comparison to the spectrum analyzer's measurement noise floor. \textbf{(b)} The cumulative phase noise in an atom interferometer as calculated with \eqref{equ:Sensitivity_PhaseNoise} and the laser phase noise PSD of (a) weighted by \eqref{equ:WeightingFunction} for a MZ pulse sequence with $\tau_\mathrm{R}=\SI{25}{\micro\second}$ and $T=\SI{120}{\milli\second}$.}
\label{fig:PhaseNoisePSDCumulative}
\end{figure}

Because the two frequency components share the same optical path in the OSSB laser system, phase noise from perturbations in the optical path up to the atom interferometer is common-mode and cancels out in the atomic state population. The phase noise at the fiber output is therefore limited by the phase noise of the RF source as observed in the measurements in {Fig.~\ref{fig:PhaseNoisePSDCumulative}}. The phase noise power spectral density (PSD) of {Fig.~\ref{fig:PhaseNoisePSDCumulative}a} is measured by a signal source analyzer (Agilent, N9030B). It is observed that the phase noise of the optical beat signal between carrier and sideband of the Raman laser is limited by the \SI{7}{\giga\hertz} signal generator source at frequencies below \SI{500}{\hertz}. Between \SI{500}{\hertz} and about \SI{100}{\kilo\hertz} the phase noise contribution from the AFG is dominating. At higher frequencies the measured phase noise of the laser beat signal is higher than that of the RF modulation signal as a result of intensity noise of the fiber laser.

The effect of the laser phase noise on the atom-interferometer phase noise $\Delta\Phi_{\phi}$ can be evaluated with the equation \cite{Cheinet2008}:
\begin{equation}
	\label{equ:Sensitivity_PhaseNoise}
	\Delta\Phi_{\phi} = \sqrt{ \int^{\infty}_{0}S_{\phi}(f)\vert H(2\pi f)\vert^{2}\mathrm{d}f } .
\end{equation}
Here is $S_{\phi}(f)$ the phase noise PSD and $H(2\pi f) = H(\omega)$ is the spectral weighting function of a Mach{\textendash}Zehnder type pulse sequence according to the formulation of \cite{Cheinet2008}: 
\begin{equation}
	\label{equ:WeightingFunction}
		\vert H(\omega)\vert^{2} = \left\vert\frac{4\Omega_\mathrm{R}\omega}{\omega^{2}-\Omega_\mathrm{R}^{2}}\right\vert^2 \sin^2\!\left(\frac{\omega(T+2\tau_\mathrm{R})}{2}\right) 
		\times\left[ \cos\!\left(\frac{\omega(T+2\tau_\mathrm{R})}{2}\right) + \frac{\Omega_\mathrm{R}}{\omega}\sin\!\left(\frac{\omega T}{2}\right) \right]^{2} ,
\end{equation}
where $\Omega_\mathrm{R}=\pi/(2\tau_\mathrm{R})$ is the Rabi frequency of the stimulated Raman transition. Applying the measured phase noise PSD of the laser beat signal in \eqref{equ:Sensitivity_PhaseNoise}, we calculate the cumulative phase noise for a MZ pulse sequence with $\tau_\mathrm{R}=\SI{25}{\micro\second}$ and $T=\SI{120}{\milli\second}$. The cumulative phase noise introduced in the atom interferometer is evaluated by taking the integration in \eqref{equ:Sensitivity_PhaseNoise} up to the frequency indicated in {Fig.~\ref{fig:PhaseNoisePSDCumulative}b}. This shows that the total phase noise contribution $\Delta\Phi_{\phi}$ of \SI{35}{\milli\radian} is limited by phase noise from the microwave source below \SI{10}{\hertz}. Bespoke microwave sources can reduce this phase noise contribution as demonstrated by other research groups \cite{Schmidt2010, Lautier2014a}, but the characterization here is currently limited by the noise floor of the spectrum analyzer.

\section{Laser frequency noise}

Frequency noise from the fiber laser introduces phase noise in stimulated Raman transitions with counter-propagating beams, due to a propagation delay of the retro-reflected Raman beam \cite{LeGouet2007}. The phase noise contribution $\Delta\Phi_{\nu}$ from this effect is estimated via
\begin{equation}
	\label{equ:Sensitivity_FrequencyNoise}
	\Delta\Phi_{\nu} = 2\pi t_\mathrm{d} \sqrt{ \int^{\infty}_{0} S_{\nu}(f)\left\vert H(2\pi f)\right\vert^{2} \mathrm{d}f } .
\end{equation}
Here is $S_{\nu}(f)$ the power spectral density of the laser frequency noise and $t_\mathrm{d}$ is the propagation delay time of the retro-reflected Raman beam. In our atom-interferometer setup the average distance between mirror and atom cloud is \SI{75}{\centi\meter}, resulting in a propagation delay of $t_\mathrm{d} = \SI{5}{\nano\second}$. 

\begin{figure}[htbp]
\centering
\includegraphics[width=0.85\linewidth]{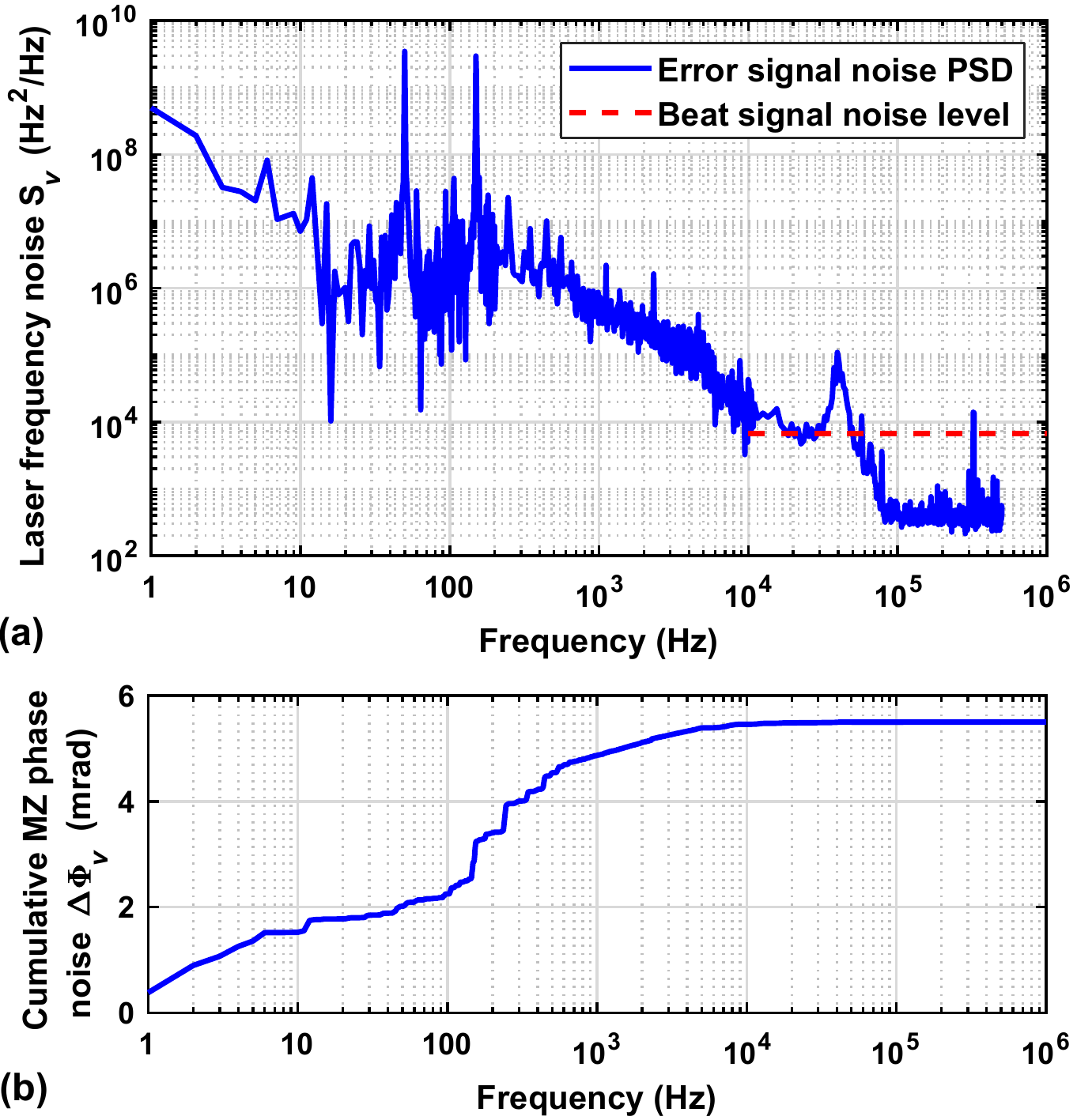}
\caption{\textbf{(a)} The power spectral density (PSD) of the laser frequency noise determined from an error signal of a rubidium modulation transfer spectroscopy setup. The PSD is averaged for frequencies above \SI{10}{\kilo\hertz} for clarity. The white noise floor level (dashed line) for frequencies above \SI{50}{\kilo\hertz} is determined from the linewidth of the beat signal spectrum with a similar laser. \textbf{(b)} The cumulative phase noise from the weighted PSD of (a) for a propagation delay of \SI{5}{\nano\second} and a MZ type pulse sequence with $\tau_\mathrm{R}=\SI{25}{\micro\second}$, $T=\SI{120}{\milli\second}$.}
\label{fig:PSDFrequencyNoise}
\end{figure}

The PSD $S_\nu(f)$ of the laser frequency noise is determined by taking the Fourier transform of the error signal from a rubidium spectroscopy setup.  A modulation transfer spectroscopy setup \cite{McCarron2008} is used to convert the optical frequency at the \SI{780}{\nano\meter} wavelength output of the laser to an error signal voltage, as described in \cite{Rammeloo2018PhD}. The fiber laser is tuned via the internal piezo element such that the laser frequency is within the linear range of the error signal around the $\ket{^{5}S_{1/2},F=2}\leftrightarrow\ket{^{5}P_{3/2},F=3}$ transition in $^{87}$Rb. In this setup no active frequency stabilization is applied, i.e. the fiber laser is unlocked. The resulting PSD of the error signal voltage is converted to laser frequency noise and plotted in {Fig.~\ref{fig:PSDFrequencyNoise}a}. The two distinct peaks visible at \SI{50}{\hertz} and \SI{150}{\hertz} in the PSD are from mains-power pick-up. The demodulation circuit of the spectroscopy setup has a cut-off frequency at \SI{50}{\kilo\hertz}, where nearby a small bump is visible in the spectrum from the resonance of the piezo tuning structure in the fiber laser.

For frequencies above \SI{50}{\kilo\hertz} the laser frequency noise is estimated by a white noise floor level of $S_\mathrm{L}=\SI{6.7e3}{\hertz\squared\per\hertz}$, as indicated by the dashed line in {Fig.~\ref{fig:PSDFrequencyNoise}a}. This white noise amplitude is calculated from the laser linewidth, which in turn is measured by creating a beat signal with a second laser of the same model. The recorded spectrum of the beat signal results in a Lorenztian laser linewidth of $\gamma=\SI{21(2)}{\kilo\hertz}$ (FWHM) and is used to estimate the laser frequency noise level via $S_\mathrm{L} = \gamma/\pi$ \cite{DiDomenico2010}. 

The cumulative phase noise in a MZ pulse sequence, plotted in {Fig.~\ref{fig:PSDFrequencyNoise}b}, is calculated from the measured PSD of {Fig.~\ref{fig:PSDFrequencyNoise}a} using \eqref{equ:WeightingFunction} and \eqref{equ:Sensitivity_FrequencyNoise}. The noise contribution in the atom-interferometry phase is found to be $\Delta\Phi_{\nu}=\SI{5.5}{\milli\radian}$ and is mostly due to the frequency noise below \SI{1}{\kilo\hertz} in the absence of active frequency stabilization. This phase noise contribution from propagation delay is similar to those reported in other laser systems \cite{LeGouet2007, Theron2014} and could also be reduced by decreasing the distance between the retro-reflecting mirror and the atom cloud.

\section{Concluding remarks}

The evaluated contributions in the atom-interferometry phase noise from the current OSSB laser system are summarized in {Table~\ref{tbl:NoiseSourcesSummary}}, together with the contribution in the relative precision or measurement bias in $g$. For comparison with other gravimeter experiments, these contributions are also evaluated for Mach{\textendash}Zehnder type pulse sequences at different interrogation times, $T=\SI{60}{\milli\second}$ and \SI{240}{\milli\second}. The most significant phase shifts come from differential light shifts due to a changing ratio between sideband and carrier intensity. Thus, for improved measurement precision and accuracy a characterization of RF phase shifts is necessary. This becomes even more crucial for increased interrogation times $T$, as seen in {Table~\ref{tbl:NoiseSourcesSummary}}, since a larger frequency chirp would be required. In comparison, the laser intensity noise does not pose a significant limit to the atom-interferometer measurement precision, because the resulting reduction in signal-to-noise ratio and additional phase shift can in principle be corrected for in post-process when monitoring the Raman beam powers and sideband/carrier ratio. We estimate that when sideband/carrier ratio fluctuations in the order of \SI{0.1}{\percent} can be achieved, the phase noise contribution $\Delta\Phi_I$ reduces to the same level as $\Delta\Phi_\nu$. Such a measurement precision of the sideband/carrier ratio is reachable with the installed Fabry{\textendash}P\'erot cavity, however the limiting factor here is its measurement speed.

The current OSSB laser system impacts our atom-interferometry experiment with $T=\SI{120}{\milli\second}$ to the level of \SI{72}{\milli\radian} per shot, corresponding to a relative measurement precision in the local gravitational acceleration of $32~\mathrm{n}g$. This study thus shows that when the sideband/carrier ratio is monitored, the performance of an OSSB laser system in atom-interferometry mainly lies in the RF design and choice of microwave components that drive the IQ modulator. The presented IQM-based laser system thus provides a tunable, phase-stable and efficient light source for high-precision atom-interferometry experiments.

\begin{table}
	\centering
	\caption{RMS phase noise and frequency chirping bias contributions from the OSSB laser system in our atom-interferometry experiment using a MZ type pulse sequences with $\tau_\mathrm{R}=\SI{25}{\micro\second}$ and different interrogation times $T$. In brackets are the corresponding values relative to the local gravitational acceleration.\label{tbl:NoiseSourcesSummary}}
	\begin{tabulary}{\linewidth}{lCCC}
		\hline
		Interrogation time $T$							&	\SI{60}{\milli\second}				&	\SI{120}{\milli\second}					&	\SI{240}{\milli\second} \\
		\hline\hline
		Laser phase noise	$\Delta\Phi_{\phi}$			& \SI{30}{\milli\radian} $(\num{52}~\mathrm{n}g)$		&	\SI{35}{\milli\radian} $(\num{16}~\mathrm{n}g)$		&	\SI{37}{\milli\radian} $(\num{4.0}~\mathrm{n}g)$	\\
		Frequency noise	$\Delta\Phi_{\nu}$	& \SI{5.0}{\milli\radian} $(\num{8.8}~\mathrm{n}g)$ &	\SI{5.5}{\milli\radian} $(\num{2.4}~\mathrm{n}g)$	&	\SI{6.1}{\milli\radian} $(\num{0.7}~\mathrm{n}g)$	\\
		Intensity noise $\Delta\Phi_I$ from:&															&																					&		\\
		- total intensity	$I_\mathrm{tot}(t)$	&							-								&	\SI{6}{\milli\radian} $(\num{2.6}~\mathrm{n}g)$		&	-	\\
		- intensity ratio $R(t)$ 						&								-								&	\SI{56}{\milli\radian} $(\num{25}~\mathrm{n}g)$		&	-	\\
		\hline
		Frequency chirping bias							& \SI{14}{\milli\radian}	$(\num{25}~\mathrm{n}g)$	&	\SI{82}{\milli\radian}	$(\num{36}~\mathrm{n}g)$	&	\SI{175}{\milli\radian}	$(\num{19}~\mathrm{n}g)$ \\
		\hline
	\end{tabulary}
\end{table}

\section{Funding}

Engineering and Physical Sciences Research Council (EP/M013294); Defence Science and Technology Laboratory (DSTLX-1000095040).

\section{Acknowledgments}

The authors thank A. Kaushik, M. Ali Khan and C. McRae for useful discussions and their help with the laser system components.

\bibliography{RammelooOSSBlasersystem}

\end{document}